\documentclass[aps, prc, floatfix, nofootinbib, superscriptaddress, twocolumn]{revtex4-1}

\usepackage{latexsym}
\usepackage{amsmath}
\usepackage{amssymb}
\usepackage{amsfonts}

\usepackage[mathscr,scaled=1.15]{urwchancal}
\DeclareFontFamily{OT1}{pzc}{}
\DeclareFontShape{OT1}{pzc}{m}{it}%
{<-> s * [1.15] pzcmi7t}{}
\DeclareMathAlphabet{\mathpzc}{OT1}{pzc}{m}{it}

\usepackage{color}

\usepackage{supertabular}
\usepackage{placeins}
\usepackage{epsfig}
\usepackage{graphicx}

\definecolor{darkgreen}{rgb}{0,0.5,0}
\definecolor{scarlet}{rgb}{1.0,0.2,0}
\definecolor{purple}{rgb}{0.5,0,0.5}
\definecolor{blue}{rgb}{0.0,0,0.9}
\definecolor{prdblue}{rgb}{0.133,0.118,0.498}
\usepackage[hypertex]{hyperref}
\hypersetup{colorlinks=true, linkcolor=purple, citecolor= purple, urlcolor=blue}

\begin{document}


\title{Parity partners in the baryon resonance spectrum}



\author{Ya Lu}
\email[]{luya@smail.nju.edu.cn}
\affiliation{Department of Physics, Nanjing University, Nanjing, Jiangsu 210093, China}

\author{Chen Chen}
\email[]{chenchen@ift.unesp.br}
\affiliation{Instituto de F\'isica Te\'orica, Universidade Estadual Paulista, Rua Dr.~Bento Teobaldo Ferraz, 271, 01140-070 S\~ao Paulo, SP, Brazil}

\author{Craig D. Roberts}
\email[]{cdroberts@anl.gov}
\affiliation{Physics Division, Argonne National Laboratory, Argonne, Illinois
60439, USA}

\author{Jorge Segovia}
\email[]{jorge.segovia@tum.de}
\affiliation{Physik-Department, Technische Universit\"at M\"unchen,
James-Franck-Str.\,1, D-85748 Garching, Germany}

\author{Shu-Sheng Xu}
\email[]{xuss@nju.edu.cn}
\affiliation{Department of Physics, Nanjing University, Nanjing, Jiangsu 210093, China}

\author{Hong-Shi Zong}
\email[]{zonghs@nju.edu.cn}
\affiliation{Department of Physics, Nanjing University, Nanjing, Jiangsu 210093, China}
\affiliation{Joint Center for Particle, Nuclear Physics and Cosmology, Nanjing, Jiangsu 210093, China}

\date{11 May 2017}

\begin{abstract}
We describe a calculation of the spectrum of flavour-$SU(3)$ octet and decuplet baryons, their parity partners, and the radial excitations of these systems, made using a symmetry-preserving treatment of a vector$\,\times\,$vector contact interaction as the foundation for the relevant few-body equations.   Dynamical chiral symmetry breaking generates nonpointlike diquarks within these baryons and hence, using the contact interaction, flavour-antitriplet scalar, pseudoscalar and vector, and flavour-sextet axial-vector quark-quark correlations can all play an active role.  The model yields reasonable masses for all systems studied, and Faddeev amplitudes for ground states and associated parity partners that sketch a realistic picture of their internal structure: ground-state, even parity baryons are constituted, almost exclusively, from like-parity diquark correlations; but orbital angular momentum plays an important role in the rest-frame wave functions of odd-parity baryons, whose Faddeev amplitudes are dominated by odd-parity diquarks.
\end{abstract}



\maketitle



\section{Introduction}\label{introduction}
%
In a symmetry-preserving treatment using relativistic quantum field theory, one may generate the interpolating field for the parity partner of any given state via a simple chiral rotation of that associated with the original state.  It follows that parity partners will be degenerate in mass and alike in structure in all theories that possess a chiral symmetry realised in the Wigner-Weyl mode.  This knowledge has long made the mass-splittings between parity partners in the strong-interaction spectrum a subject of particular interest.  A classic example is provided by the $\rho(770)$- and $a_1(1260)$-mesons: viewed as chiral and hence parity partners, it has been argued \cite{Weinberg:1967kj} that their mass and structural differences can be attributed entirely to dynamical chiral symmetry breaking (DCSB), \emph{viz}.\ realisation of chiral symmetry in the Nambu-Goldstone mode.

Potentially connected intimately with confinement \cite{Roberts:2016vyn}, DCSB is a key emergent strong interaction phenomenon; and regarding its role in explaining the splitting between parity partners, additional insights can be developed by studying the quantum field theory bound-state equations appropriate to the $\rho$- and $a_1$-mesons.  One finds that the rest-frame projections of their Poincar\'e-covariant wave functions are primarily $S$-wave in nature \cite{Maris:1999nt, Chang:2008sp, Chang:2011ei, Roberts:2011cf, Chen:2012qr, Eichmann:2016yit}, even though both possess nonzero angular momentum \cite{Bloch:1999vka, Gao:2014bca}, whose magnitude influences the size of the splitting \cite{Chang:2011ei}.

On the other hand, the picture is quite different when drawn using a quantum mechanical constituent-quark model.  Such a framework does not support the notion of chiral partners; and the $a_1$-meson is typically described as an $L=1$ orbital-angular-momentum excitation of an $L=0$ $\rho$-meson quark+antiquark system, with the roughly $500\,$MeV mass-splitting produced by tuning the model's potential \cite{DeRujula:1975qlm, Godfrey:1985xj}.  Such conflicts of interpretation are repeatedly encountered in the spectrum of light-quark mesons.

They are also common in the baryon spectrum.  For instance, quark models predict that the $N(1535)\tfrac{1}{2}^-$, $\Lambda(1670)\tfrac{1}{2}^-$, $\Sigma(1750)\tfrac{1}{2}^-$ baryons are all states with three constituent-quarks possessing one unit of orbital angular momentum, $L=1$ \cite{Olive:2016xmw}.  However, in quantum field theory these systems appear as chiral partners of the ground state baryons $N(939)\tfrac{1}{2}^+$, $\Lambda(1116)\tfrac{1}{2}^+$, $\Sigma(1193)\tfrac{1}{2}^+$, in which case, again, the roughly $500\,$MeV mass differences should chiefly be generated by DCSB.  In this connection, elucidating the details of the mechanism would be of great interest and value.

The preceding discussion highlights that baryon spectroscopy has played a key role in developing our understanding of strong-interaction dynamics and formulating quantum chromodynamics (QCD) \cite{Aznauryan:2011qj, Aznauryan:2012baS, Briscoe:2015qia}, and yet quark-model concepts still seem a sound way to label and organise the spectrum of baryon resonances.  It is therefore worth exploring the field theory perspective, in search of both common positions and explanations for apparent conflicts.  Herein, we pursue this by employing quantum field equations in the continuum and computing the spectrum produced by a particularly simple interaction, which enables an algebraic analysis of the issues involved.
The interaction, the diquark correlations it supports, and the baryon bound-state equations are all explained and defined in Sec.\,\ref{sectiontwo}; our results are presented and discussed in Sec.\,\ref{sectionthree}; and Sec.\,\ref{sectionconclusion} summarises our findings and provides them with a context.


\section{Bound-State Equations}\label{sectiontwo}
%
\subsection{Quark-quark interaction}
The Dyson-Schwinger equations \cite{Roberts:2015lja, Aguilar:2015bud} provide a natural framework for the symmetry-preserving treatment of hadron bound states in quantum field theory; and the starting point in the matter sector is knowledge of the quark-quark interaction.  This is now known with some certainty \cite{Binosi:2014aea, Binosi:2015xqk, Binosi:2016wcx, Binosi:2016xxu, Binosi:2016nme, Aguilar:2002tc, Brodsky:2008be, Deur:2016tte}, as are its consequences: whilst the effective charge, and gluon and quark masses run with momentum, $k^2$, they all saturate at infrared momenta, each changing by $\lesssim20$\% on $0\lesssim \surd k^2 \lesssim m_g \approx m_p/2$, where $m_g$ is a renormalisation-group-invariant gluon mass-scale and $m_p$ is the proton mass.  It follows that, employed judiciously, a symmetry-preserving treatment of a vector$\,\times\,$vector contact interaction can provide insights and useful results for those hadron observables whose measurement involves probe momenta less-than $m_g$ \cite{Farias:2005cr, Farias:2006cs, GutierrezGuerrero:2010md, Roberts:2010rn, Roberts:2011cf, Roberts:2011wy, Wilson:2011aa, Chen:2012qr, Pitschmann:2012by, Chen:2012txa, Wang:2013wk, Segovia:2013rca, Segovia:2013uga, Pitschmann:2014jxa, Xu:2015kta, Bedolla:2015mpa, Bedolla:2016yxq, Serna:2016ifh}.  Hadron masses are such quantities.

\begin{table}[t]
\caption{\label{tabledressedquark}
Computed dressed-quark properties, required as input for the bound-state equations employed herein.  All results obtained with contact-interaction strength $\alpha_{\rm IR} =0.93 \pi$, and (in GeV) infrared and ultraviolet regularisation scales $\Lambda_{\rm ir} = 0.24=1/r_{\rm ir}$, $\Lambda_{\rm uv}=0.905=1/r_{\rm uv}$, respectively.  N.B.\ These parameters take the values determined in the spectrum calculation of Ref.\,\protect\cite{Chen:2012txa}, we assume isospin symmetry throughout, and $\Lambda_{\rm ir}>0$ implements dressed-quark confinement \cite{Ebert:1996vx}.
(All dimensioned quantities are listed in GeV.)}
\begin{center}
\begin{tabular*}
{\hsize}
{
c@{\extracolsep{0ptplus1fil}}
c@{\extracolsep{0ptplus1fil}}
c@{\extracolsep{0ptplus1fil}}
c@{\extracolsep{0ptplus1fil}}
c@{\extracolsep{0ptplus1fil}}
c@{\extracolsep{0ptplus1fil}}
c@{\extracolsep{0ptplus1fil}}
c@{\extracolsep{0ptplus1fil}}}\hline
\multicolumn{4}{c}{input: current masses} & \multicolumn{4}{c}{output: dressed masses}\\
$m_0$ & $m_u$ & $m_s$ & $m_s/m_u$ & $M_0$ &   $M_u$ & $M_s$ & $M_s/M_u$  \\\hline
0 & 0.007  & 0.17 & 24.3 & 0.36 & 0.37 & 0.53 & 1.43  \\\hline
\end{tabular*}
\end{center}
\end{table}

In the context of a vector$\,\times\,$vector contact interaction, a symmetry-preserving formulation of the coupled two- and three-valence-body bound-state problems is detailed in Refs.\,\cite{Roberts:2011cf, Chen:2012qr}.  It is based upon the rainbow-ladder (RL) approximation, which is the leading-order in a systematic DSE truncation scheme \cite{Binosi:2016rxz}, and we follow that approach herein, using the same parameter values.  They are reported in Table~\ref{tabledressedquark}, along with the results they yield for the masses of the dressed $u=d$- and $s$-quarks when used in the rainbow-truncation gap equation.

\begin{figure}[t]
\centerline{%
\includegraphics[clip,width=0.45\textwidth]{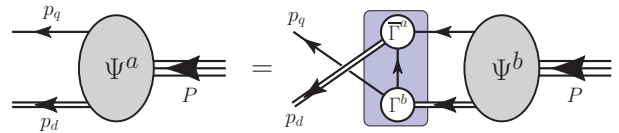}}
\caption{\label{figFaddeev} Poincar\'e covariant Faddeev equation.  $\Psi$ is the Faddeev amplitude for a baryon of total momentum $P= p_q + p_d$.  The shaded rectangle demarcates the kernel of the Faddeev equation: \emph{single line}, dressed-quark propagator; $\Gamma$,  diquark correlation amplitude; and \emph{double line}, diquark propagator.}
\end{figure}

In this approach, baryon masses are obtained by solving a Poincar\'e-covariant Faddeev equation \cite{Cahill:1988dx, Burden:1988dt, Cahill:1988zi, Reinhardt:1989rw}, which sums all possible quantum field theoretical exchanges and interactions that can take place between the three dressed-quarks that characterise its valence-quark content.  A dynamical prediction of Faddeev equation studies is the appearance of nonpointlike, colour-antitriplet quark$+$quark (diquark) correlations within baryons, whose characteristics are greatly influenced by DCSB \cite{Cahill:1987qr, Bloch:1999vk, Maris:2002yu, Wang:2005tq, Wang:2011ab, Segovia:2015ufa, Santopinto:2016fay}.  Consequently, the baryon bound-state problem is transformed into the exercise of solving the linear, homogeneous matrix equation depicted in Fig.\,\ref{figFaddeev}.

\subsection{Diquark correlations}
As highlighted by Fig.\,\ref{figFaddeev}, in order to complete the Faddeev equation kernels for all octet and decuplet baryons, and their parity partners, it is necessary to compute the masses and canonically normalised correlation amplitudes for each diquark system that can contribute to these bound states.  This is readily achieved by solving Bethe-Salpeter equations in all relevant channels, once those channels are identified.

Regarding colour, all participating diquark correlations are antitriplets because they must combine with the bystander quark to form a colour singlet.  Notably, the colour-sextet quark+quark channel does not support correlations because gluon exchange is repulsive in this channel \cite{Cahill:1987qr}.

Diquark spin-flavour structure is more complex.  It is revealed by solving the Bethe-Salpeter equations, with the result that a contact interaction supports the following systems (the subscript indicates $J^P$):
\begin{subequations}
\label{qqAmplitudes}
\begin{align}
\Gamma_{0^+}^j(K) & = \vec{H} \, T_{\bar 3_f}^j [i\gamma_5 E^j_{0^+} + \frac{1}{M_R^j} \gamma_5\gamma\cdot K F^j_{0^+}] \, C\,,\\
\Gamma_{1^+\mu }^f(K) & = \vec{H} \, T_{6_f}^g \,\tilde\gamma_\mu C \, E_{1^+}^f\,,\\
\Gamma_{0^-}^j(K) & =\vec{H}  \, T_{\bar 3_f}^j \, i C \,E^j_{0^-} \\
\Gamma_{1^-\mu }^j(K) & = \vec{H} \, T_{\bar 3_f}^j \,\tilde\gamma_\mu \gamma_5 C \,E_{1^-}^j\,,
\end{align}
\end{subequations}
where $K$ is the total momentum of the correlation,
$K\cdot \tilde\gamma=0$;
$M_R^{1}=M_u=M_d$, $M_R^{2,3}=2 M_u M_s/(M_u+M_s)$;
$\vec{H} = \{i\lambda_c^7, -i\lambda_c^5,i\lambda_c^2\}$, with $\{\lambda_c^k,k=1,\ldots,8\}$ denoting Gell-Mann matrices in colour space, expresses the diquarks' colour antitriplet character;
$C$ is the charge-conjugation matrix;
\begin{subequations}
\label{Tmatrices}
\begin{align}
\{T_{\bar 3_f}^j,j& =1,2,3\} = \{i\lambda^2,i\lambda^5,i\lambda^7\}\,, \\
\nonumber
\{T_{6_f}^g,g & =1,\ldots,6\}  =\{
s_0 \lambda^0 +s_3 \lambda^3  + s_8 \lambda^8 ,
\lambda^1,
\lambda^4, \\
&  s_0 \lambda^0 - s_3 \lambda^3  + s_8 \lambda^8,
\lambda^6,
s_0 \lambda^0 - 2 s_8 \lambda^8
\}\,,
\end{align}
\end{subequations}
with $s_0=\surd 2/3$, $s_3=1/\surd 2$, $s_8=1/\surd 6$,
$\{\lambda^k,k=1,\ldots,8\}$ denoting flavour Gell-Mann matrices,
$\lambda^0={\rm diag}[1,1,1]$, and
all flavour matrices left-active on column$[u,d,s]$.
Notably, using a contact interaction, the amplitudes do not depend on the relative momentum, \emph{i.e}.\ angular momentum is suppressed in these bound-states; and it follows that a flavour-sextet vector-diquark is not supported by a RL-like treatment of the contact interaction \cite{Roberts:2011wy}.  (Our Euclidean metric conventions are explained in Appendix\,A of Ref.\,\cite{Chen:2012qr}.)

We obtain the masses and amplitudes of the contact-interaction diquark correlations by solving the appropriate Bethe-Salpeter equations in a RL-like truncation.  (At this level, the equation for a $J^{P}$ diquark is readily obtained from that for a $J^{-P}$ meson \cite{Cahill:1987qr}.)  However, the pseudoscalar and vector diquark equations are adjusted in order to implement DCSB effects that are crucial for a successful description of the meson spectrum but only appear in truncations that nonperturbatively improve upon RL \cite{Chang:2011ei}.  These adjustments mock-up the impact of DCSB-enhanced ``spin-orbit'' repulsion in the Bethe-Salpeter kernels that feed upon $L\neq 0$ components of meson wave functions, which are normally subleading in light-quark two-body systems.  In Refs.\,\cite{Roberts:2011cf, Chen:2012qr}, the desired result was achieved by including a multiplicative factor $g_{\rm SO}=0.24$ in the Bethe-Salpeter kernels for the $0^-$, $1^-$ diquarks: the value was chosen to give the empirical size of the $a_1$-$\rho$ mass-splitting.  (N.B.\,In a Poincar\'e covariant treatment, no bound-state can be purely $S$- or $P$-wave; and $g_{\rm SO}=1$ means no additional repulsion beyond that generated by the RL kernel.)

In extending a contact-interaction analysis of the spectrum and structure of baryons to include opposite-parity diquark correlations, \emph{e.g}.\ both $P=+$ and $P=-$ diquarks in $P=+$ baryons whenever physically allowed, we find the approach of Refs.\,\cite{Roberts:2011cf, Chen:2012qr} to be inadequate because it produces a splitting between $1^-$ and $0^-$ diquarks that is too small to support a realistic spectrum.  For instance, one finds that the parity-partner of the $\Xi$-baryon is unbound.  The first step toward eliminating this weakness is made by introducing two distinct SO-parameters in the meson sector, one for the axial-vector channel and another for scalar: with
\begin{equation}
\label{eqgSO}
g_{\rm SO}^{1^+_{q\bar q}} = 0.25\,, \; g_{\rm SO}^{0^+_{q\bar q}} = 0.32\,,
\end{equation}
one obtains $m_{a_1}-m_\rho = 0.45\,$GeV, in line with experiment, and $m_{\sigma_{\rm core}} - m_\rho = 0.29\,$GeV, which matches the splitting produced by the sophisticated Bethe-Salpeter kernels described in Ref.\,\cite{Chang:2011ei}.  The spectrum of ground-state mesons thus obtained is listed in Table~\ref{Diquarkmasses}.

\begin{table}[t]
\caption{\label{Diquarkmasses}
\emph{Upper panel}.  Quark-core masses of ground-state mesons computed using a symmetry-preserving regularisation of the vector$\times$vector contact interaction, with the input from Table~\protect\ref{tabledressedquark} and using Eq.\,\eqref{eqgSO}.
\emph{Middle panel}.
Row~1: Mass-scales associated with diquark correlations that play a role in the octet and decuplet spectra of baryons, computed as described in connection with Eqs.\,\eqref{eqgSO}, \eqref{eqsSO}.
Rows~2 and 3: Canonically normalised, momentum-independent Bethe-Salpeter amplitudes associated with each diquark, Eqs.\,\eqref{qqAmplitudes}.
\emph{Lower panel}.
Analogous results from Ref.\,\cite{Chen:2012qr} for comparison.
Notably, in all cases the correlation mass increases by $\approx 0.1\,$GeV with the addition of each dressed $s$-quark.
(All dimensioned quantities are listed in GeV.)
}
\begin{center}
\begin{tabular*}
{\hsize}
{
l@{\extracolsep{0ptplus1fil}}
|l@{\extracolsep{0ptplus1fil}}
l@{\extracolsep{0ptplus1fil}}
|l@{\extracolsep{0ptplus1fil}}
l@{\extracolsep{0ptplus1fil}}
l@{\extracolsep{0ptplus1fil}}
|l@{\extracolsep{0ptplus1fil}}
l@{\extracolsep{0ptplus1fil}}
|l@{\extracolsep{0ptplus1fil}}
l@{\extracolsep{0ptplus1fil}}}\hline
 $\rule{1em}{0ex}$ & $\pi$ & $K$ &
$\rho$ & $K^\ast $ & $\phi\;\;$ &
$\sigma$ & $\kappa$ &
$a_1$ & $K_1$  \\\hline
m &
0.14 & 0.50 &
0.93 & 1.03 & 1.13 &
1.22 & 1.32 &
1.37 & 1.48 \\\hline\hline
\multicolumn{10}{c}{  }\\\hline
 $\rule{1em}{0ex}$ & $0^{+(1)}_{\bar 3_f}$ & $0^{+(2,3)}_{\bar 3_f}$ &
$1^{+(1,2,4)}_{6_f}$ & $1^{+(3,5)}_{6_f}$ & $1^{+6}_{6_f}\;\;$ &
$0^{-(1)}_{\bar 3_f}$ & $0^{-(2,3)}_{\bar 3_f}$ &
$1^{-(1)}_{\bar 3_f}$ & $1^{-(2,3)}_{\bar 3_f}$  \\\hline
m &
0.78 & 0.93 &
1.06 & 1.16 & 1.26 &
1.15 & 1.26 &
1.33 & 1.44 \\\hline
E &
2.74 & 2.88 &
1.30 & 1.36 & 1.42 &
1.06 & 1.08 &
0.51 & 0.50 \\\hline
F &
0.31 & 0.39 &
 &  &  &
 &  &
 &   \\\hline\hline
%
%
\multicolumn{10}{c}{Ref.\,\cite{Chen:2012qr} }\\\hline
m &
0.78 & 0.93 &
1.06 & 1.16 & 1.26 &
1.37 & 1.47 &
1.45 & 1.55 \\\hline
E &
2.74 & 2.91 &
1.30 & 1.36 & 1.42 &
0.40 & 0.39 &
0.27 & 0.27 \\\hline
F &
0.31 & 0.40 &
 &  &  &
 &  &
 &   \\\hline\hline
\end{tabular*}
\end{center}
\end{table}

We would like to emphasise here that the meson labels indicate the quark-cores of the respective systems, which are distinguished in a well-defined manner from the empirical states, as detailed, \emph{e.g}.\ in Refs.\,\cite{Roberts:2011cf, Chen:2012qr}.
To expand a little, owing to our choice for the current-quark masses, $m_\pi$ and $m_K$ agree with experiment, but all other computed values for ground-state masses are greater than the empirical values.  This is typical of DCSB-corrected kernels that nevertheless omit resonant contributions, \emph{i.e}.\ do not contain effects that may be associated with a meson cloud.  Such kernels should produce dressed-quark-core masses for hadron ground-states that are larger than the empirical values, thereby leaving room for resonant corrections to reduce the mass \cite{Roberts:1988yz, Hollenberg:1992nj, Alkofer:1993gu, Mitchell:1996dn, Ishii:1998tw, Pichowsky:1999mu, Hecht:2002ej, Eichmann:2008ae}.  For instance, the most complete computation of this sort predicts that such effects reduce $m_\rho$ by $0.13\,$GeV \cite{Pichowsky:1999mu}.  Applied to our result, this would produce $m_\rho^{\rm loop-corrected} = 0.8\,$GeV, in fair agreement with the empirical value: $0.78\,$GeV.

The second step is to dampen spin-orbit effects in diquark channels, \emph{viz}.\ we write:
\begin{equation}
\label{eqsSO}
g_{\rm SO}^{1^-_{qq},0^-_{qq}} = g_{\rm SO}^{1^+_{q\bar q},0^+_{q\bar q}} \times {\mathpzc s}_{\rm SO} \,,\;
{\mathpzc s}_{\rm SO}=1.8\,,
\end{equation}
so that the modification factor in our RL-like diquark Bethe-Salpeter equations is nearer unity and hence generates less repulsion.  Physically, this might be understood by acknowledging that valence-quarks within a diquark are more loosely correlated than the valence-quark and -antiquark pair in a bound-state meson and, consequently, spin-orbit repulsion in diquarks should be less pronounced than it is in mesons.
The spectrum of diquark mass-scales, obtained following the procedure described above, is listed in Table~\ref{Diquarkmasses}: a $\pm 10$\% change in ${\mathpzc s}_{\rm SO}$ alters the $0^-$, $1^-$ values by $\mp 2$\%.

The size of ${\mathpzc s}_{\rm SO}$ in Eq.\,\eqref{eqsSO} is explained below.  Meanwhile, inspection of Table~\ref{Diquarkmasses} reveals that this value serves to reverse the $0_{q\bar q}^+-0_{qq}^-$, $1_{q\bar q}^+-1_{qq}^-$ meson-diquark mass orderings that are characteristic of RL truncation.  As we shall see, such a reversal in the $0^-$, $1^-$ diquark channels seems to be required if one wishes to obtain a realistic baryon spectrum from the contact interaction.  Whether such ordering is itself realistic, however, ought to be checked using the most sophisticated Bethe-Salpeter kernels that are currently available.

The truncation employed herein generates asymptotic (freely propagating) diquarks.  Such states are not observed empirically and their appearance is an artefact of the truncation.  Higher-order terms in the quark-quark scattering kernel, whose analogue in the $q \bar q$-channel do not materially affect meson properties, ensure that QCD's quark-quark scattering matrix does not exhibit singularities which correspond to asymptotic diquark states \cite{Bender:1996bb, Bhagwat:2004hn}.  Studies with kernels that exclude diquark bound states nevertheless support a physical interpretation of the diquark mass-scales, $m_{(qq)_{\!J^P}}$, obtained using our truncation, \emph{viz}.\ the quantity $\ell_{(qq)^{\!J^P}}:=1/m_{(qq)_{\!J^P}}$ may be interpreted as a range over which the diquark correlation can propagate before fragmentation.

\subsection{Faddeev equations}
Fig.\,\ref{figFaddeev} shows that the Faddeev kernels involve diquark breakup and reformation via exchange of a dressed-quark.  In order to present the most transparent analysis possible, we follow Refs.\,\cite{Roberts:2011cf, Chen:2012qr} and introduce a simplification, \emph{viz}.\ in the Faddeev equation for a baryon of type $B$, the quark exchanged between the diquarks is represented as
\begin{equation}
S_h(k) \to \frac{g_B^2}{M_h}\,,
\label{staticexchange}
\end{equation}
where $h=u,d,s$ is the quark's flavour and $g_B$ is a coupling constant.  This is a variant of the  ``static approximation,'' which itself was introduced in Ref.\,\cite{Buck:1992wz}.  It has a marked impact on the Faddeev amplitudes, forcing them to be momentum-independent, just like the diquark Bethe-Salpeter amplitudes, but calculations reveal that it has little impact on the computed masses \cite{Xu:2015kta}.  The values
$g_{8\equiv{\rm octet}}=1.18$, 
$g_{10\equiv{\rm decuplet}}=1.56$,
were fixed in Ref.\,\cite{Roberts:2011cf} in order to produce masses for the nucleon and $\Delta$-baryon that are each inflated by roughly 0.2\,GeV in order to ensure that the experimental values are reproduced after meson-baryon final-state interactions are incorporated \cite{Ishii:1998tw, Hecht:2002ej, Eichmann:2008ae, Eichmann:2008ef, Suzuki:2009nj, Kamano:2013iva, Segovia:2014aza, Roberts:2015dea, Segovia:2015hra}.

With the inputs described above one can construct all Faddeev kernels associated with ground-state octet and decuplet baryons, and their parity partners.  The structure of those kernels depends on the form of the baryon Faddeev amplitudes.  They are simplest for the decuplet baryons because their spin-flavour composition precludes a role for flavour-$\bar 3$ diquark correlations.
For example, the amplitudes for the positive-energy, doubly-positive electric-charge $\Delta_{P=\pm}$ baryons are $\Psi_\mu^\pm = \psi_{\mu\nu}^\pm (P)u_\nu(P)$, where $u_\nu(P)$ is a Rarita-Schwinger spinor, $P$ is the baryon's total momentum, and
\begin{subequations}
\label{DeltaAmp}
\begin{align}
\psi_{\mu\nu}^\pm (P) &  u_\nu(P)  =
\Gamma^{1}_{1^+\mu } \Delta^{1^+}_{\mu\nu}(K) {\mathpzc D}_{\nu\rho}^{\pm 1}(P) u_\rho(P) \,,\\
{\mathpzc D}_{\nu\rho}^{\pm 1} & = f^\pm \delta_{\nu\rho}\mathpzc{G}^\pm\,, \\
\label{qq1prop}
\Delta^{1^+}_{\mu\nu} & =[\delta_{\mu\nu} + K_\mu K_\nu/m_{\{uu\}}^2]/[K^2+ m_{\{uu\}}^2]\,,
\end{align}
\end{subequations}
with $m_{\{uu\}}= m_{1^+}^1$ ($=1.06\,$GeV from Table~\ref{Diquarkmasses}),
$\mathpzc{G}^{+(-)} = \mathbf{I}_{\rm D} (\gamma_5)$.
(In the isospin symmetric limit, all charge-states of the $\Delta$-baryons are degenerate.)

Using Eqs.\,\eqref{DeltaAmp} and following the procedure detailed in Refs.\,\cite{Roberts:2011cf, Chen:2012qr}, one readily arrives at the following one-dimensional eigenvalue equations for the masses $m_{\Delta_\pm}$:
\begin{align}
\nonumber
1& = \frac{g_{10}^2}{2\pi^2 M_u} \frac{E_{\{uu\}}^2}{m_{\{uu\}}^2}
\int_0^1 \! d\alpha \, \overline{\mathpzc{C}}_1^{\rm iu}(\sigma) \\
&  \times [ m_{\{uu\}}^2 + (1-\alpha)^2 m_{\Delta_\pm}^2 ] [M_u \pm \alpha m_{\Delta_\pm}]\,,
\label{DeltaFE}
\end{align}
where $E_{\{uu\}} = E_{1^+}^{1}$,
\begin{equation}
\label{barC1}
\overline{\mathpzc{C}}_1(\sigma) = \Gamma(0,\sigma r_{\rm uv}^2) - \Gamma(0,\sigma r_{\rm ir}^2)\,,
\end{equation}
with $\Gamma(\alpha,y)$ being the incomplete gamma-function, and $\sigma=(1-\alpha) M_u^2 + \alpha m_{\{uu\}}^2 -\alpha(1-\alpha) m_{\Delta_\pm}^2$.  (N.B.\ Refs.\,\cite{Roberts:2011cf, Chen:2012qr} assumed that the parity partner of a given baryon is obtained by replacing the diquark correlation(s) involved by its (their) parity partners.  This overlooked the fact that a RL treatment of the contact-interaction does not support flavour-sextet vector diquarks.  The $\Delta_-$ equation given here is therefore the preferred form.)

The Faddeev amplitudes for the positive-energy nucleon and its parity partner, $\Psi^\pm = \psi^\pm u(P)$, are:
{\allowdisplaybreaks
\begin{align}
\nonumber
\psi^\pm u(P)  & =  \Gamma^1_{0^+} \Delta^{0^+}(K)\, {\mathpzc S}^\pm(P)u(P)   \\
\nonumber
& + \mbox{$\sum$}_{f=1,2}\Gamma^{f}_{1^+\mu }\Delta^{1^+}_{\mu\nu}(K) {\mathpzc A}_\nu^{\pm f}(P) u(P)\\ %
\nonumber
& + \Gamma^1_{0^-}(K) \Delta^{0^-}(K) {\mathpzc P}^{\pm}(P)\,u(P)  \\
& + \Gamma^{1}_{1^-\mu } \Delta^{1^-}_{\mu\nu}(K) {\mathpzc V}^\pm_\nu(P) u(P)\,,
\label{nucleonamplitude}
\end{align}}
\hspace*{-0.6\parindent}where $u(P)$ is a Dirac spinor;
$\Delta^{0^+}(K)$, etc.\ are standard propagators for scalar or vector bosons, detailed in Refs.\,\cite{Roberts:2011cf, Chen:2012qr}, with the appropriate masses from Table~\ref{Diquarkmasses}, analogous to Eq.\,\eqref{qq1prop};
and, with $\hat P^2=-1$,
\begin{align}
\nonumber
{\mathpzc S}^\pm & = \mathpzc{s}^\pm\,\mathbf{I}_{\rm D} \mathpzc{G}^\pm \,, \quad
i {\mathpzc P}^\pm  = \mathpzc{p}^\pm\,  \gamma_5 \mathpzc{G}^\pm\,,
\\
i{\mathpzc A}_\mu^{\pm f} & = (\mathpzc{a}_1^{\pm f} \gamma_5\gamma_\mu - i \mathpzc{a}_2^{\pm f} \gamma_5 \hat P_\mu) \mathpzc{G}^\pm \,, \label{spav} \\
\nonumber
i \mathpzc{V}_\mu^\pm & = (\mathpzc{v}_1^\pm \gamma_\mu - i \mathpzc{v}_2^\pm \mathbf{I}_{\rm D}  \hat P_\mu)\gamma_5\mathpzc{G}^\pm\,.
\end{align}

In order to obtain the masses, $m^2_\pm$, and eigenvectors $(\mathpzc{s}^\pm,\mathpzc{a}_1^{\pm f},\mathpzc{a}_2^{\pm f},\mathpzc{p}^\pm,\mathpzc{v}_1^{\pm}, \mathpzc{v}_2^{\pm})$, one substitutes the amplitudes from Eq.\,\eqref{nucleonamplitude} into the Faddeev equation depicted in Fig.\,\ref{figFaddeev} and solves the resulting eigenvalue problems.  The explicit form of the nucleon's Faddeev equation, obtained in the absence of pseudoscalar and vector diquarks, is derived in Refs.\,\cite{Roberts:2011cf, Chen:2012qr}.  Owing to isospin symmetry, the kernel can be reduced to a $3\times 3$ matrix because $\mathpzc{a}_{\{ud\}}^{\pm} = - \mathpzc{a}_{\{uu\}}^{\pm}/\surd 2$.  The extended equation, including pseudoscalar and vector diquarks, and the analogous complete equation for the nucleon's parity partner can both be obtained by following the same procedure.  The kernels may be expressed as  $6\times 6$ matrices, each element of which has the type of algebraic structure expressed in Eq.\,\eqref{DeltaFE}.  (We list the spin-flavour composition of all baryon Faddeev amplitudes in Eqs.\,\eqref{FaddeevSpinFlavour}.)

The Faddeev equations for the other octet baryons and their parity partners can similarly be obtained.
In fact, if one keeps track of isospin, then the $\Sigma_\pm$ equations may simply be obtained from the $N_\pm$ equations by replacing the $d$-quark by the $s$ quark, and the $\Xi_\pm$ equations by making the exchange $s\leftrightarrow u$ in the $\Sigma_\pm$ equations.
Regarding the $I=0$ $\Lambda_\pm$-baryons, one must pay some attention to ensuring the correct flavour structure, but this is made straightforward by following Ref.\,\cite{Chen:2012qr}.  The procedure yields an $8\times 8$-matrix Faddeev kernel describing systems that involve only the $I=0$ combinations of diquarks identified in Eq.\,\eqref{flavourLambda}.

Solving the equations thus obtained, one finds that the ground-state $P=+$ octet baryons are primarily constituted from like-parity diquarks, with negligible contributions from $P=-$ correlations.  This was to be anticipated, given the quality of existing studies that omitted $P=-$ diquarks \cite{Roberts:2011cf, Chen:2012qr}.  Unexpectedly, on the other hand, the parity partners of the ground-state octet baryons: $N^\ast(1535)\tfrac{1}{2}^-$, $\Lambda(1670)\tfrac{1}{2}^-$, $\Sigma(1750)\tfrac{1}{2}^-$, $\Xi\tfrac{1}{2}^-$, are also dominated by $P=+$ diquark correlations and, consequently, too light.  It appears, therefore, that something important is missing from the RL-like contact-interaction treatment of odd-parity octet baryons.  As with mesons and diquarks, the obvious candidate is spin-orbit repulsion.  We therefore introduce a new parameter into the Faddeev equation for $J^P=(1/2)^P$ baryons: $g_{\rm DB}$, a linear multiplicative factor attached to each unequal-parity diquark amplitude in the baryon's Faddeev equation kernel.  For example, in the nucleon's Faddeev equation:
\begin{equation}
\label{eqgDB1}
N(940) \left|
\begin{array}{ll}
E_{0^+} E_{0^+} & \to \, E_{0^+} E_{0^+} \\
E_{0^+} E_{0^-}  &\to g_{\rm DB} \, E_{0^+} E_{0^-} \\
E_{0^-} E_{0^-}  &\to g_{\rm DB}^2 \, E_{0^-} E_{0^-}
\end{array}\right.\,,
\end{equation}
with like treatment of all similar terms; whereas in the equation for the nucleon's parity partner, the pattern is reversed:
\begin{equation}
\label{eqgDB2}
N^\ast(1535) \left|
\begin{array}{ll}
E_{0^+} E_{0^+} & \to g_{\rm DB}^2 \, E_{0^+} E_{0^+} \\
E_{0^+} E_{0^-}  &\to g_{\rm DB} \, E_{0^+} E_{0^-} \\
E_{0^-} E_{0^-}  &\to E_{0^-} E_{0^-}
\end{array}\right. \,.
\end{equation}

The contact-interaction Faddeev equations for decuplet baryons are very simple, with both positive and negative parity systems involving only $6_f$ axial-vector diquarks.  Consequently, we do not include a similar factor, thereby assuming implicitly that spin-orbit repulsion is less-important, but not unimportant, in these $J=(3/2)$ states.  Whether this contact-interaction feature is sensible, or not, can only be tested by using more realistic Faddeev kernels.  Some studies exist \cite{Eichmann:2016nsu}, but it is desirable to conduct such analyses using kernels that leave room for contributions from meson-baryon final-state interactions and are capable of yielding testable resonance structure predictions on a large domain of spacelike momenta \cite{Segovia:2014aza, Roberts:2015dea, Segovia:2015hra}.

Faddeev equations describing all ground-state octet and decuplet baryons, and their parity partners are now in hand.  At this point it is natural to ask whether one can simultaneously address the radial excitations of these systems.
In quantum mechanics, the radial wave function for a bound-state's first radial excitation possesses a single zero.  A similar feature is expressed in quantum field theory: namely, in a fully covariant approach, a single zero is usually seen in the relative-momentum dependence of the leading Chebyshev moment of the dominant Dirac term in the bound state amplitude for a hadron's first radial excitation \cite{Holl:2004fr, Qin:2011xq, Segovia:2015hra}.  The existence of radial excitations is therefore clear evidence against the possibility that the interaction between quarks is momentum-independent: a bound-state amplitude that is independent of the relative momentum cannot exhibit a zero.  One may also express this differently, \emph{viz}.\ if the location of the zero is at $k_0^2$, then it is only possible for a momentum-independent interaction to produce reasonable results for phenomena that probe momentum scales $k^2\ll k_0^2$.  Typically, $k_0^2 \sim 2 M^2 \sim (0.5\,{\rm GeV})^2$ \cite{Holl:2004fr, Qin:2011xq,  Segovia:2015hra}.

\begin{table*}[t]
\caption{\label{OctetDecupletMasses}
Dressed-quark-core masses of ground-state octet and decuplet baryons, those of their radial excitations and of all their parity partners, computed using the formulation of a vector$\times$vector contact interaction described herein.
Row~1: Baryon ground-states.  The lowest mass state in each channel has positive parity.
Row~3: First radial excitations of the ground-states, which provide the second level in each channel.
Row~5: Parity partners of the baryon ground-states, which provide the third level in each channel.
Row~7: First radial excitations of the parity partner to each of the baryon ground-states, which provide the fourth level in each channel.
Masses in the rows labelled ``expt.'' are taken from Ref.\,\protect\cite{Olive:2016xmw}: where the estimated uncertainty in the location of a resonance's pole position is noticeable, it is indicated by an error bar in Fig.\,\ref{PlotOctetDecupletMasses}; and a hyphen in any position indicates that no empirically known resonance can confidently be associated with the theoretically predicted state.
(All dimensioned quantities are listed in GeV.)
}
\begin{center}
\begin{tabular*}
{\hsize}
{
l@{\extracolsep{0ptplus1fil}}
l@{\extracolsep{0ptplus1fil}}
l@{\extracolsep{0ptplus1fil}}
|l@{\extracolsep{0ptplus1fil}}
l@{\extracolsep{0ptplus1fil}}
l@{\extracolsep{0ptplus1fil}}
l@{\extracolsep{0ptplus1fil}}
|l@{\extracolsep{0ptplus1fil}}
l@{\extracolsep{0ptplus1fil}}
l@{\extracolsep{0ptplus1fil}}
l@{\extracolsep{0ptplus1fil}}}\hline
& & & \rule{0ex}{2.5ex}
$N$ & $\Lambda$ & $\Sigma$ & $\Xi$
   & $\Delta$ & ${\Sigma^\ast}$ & ${\Xi^\ast}$ & $\Omega$ \\\hline
$P=+$ & n=0 & \rule{0ex}{2.5ex}
DSE & 1.14 & 1.26 & 1.36 & 1.43 & 1.39 & 1.51 & 1.63 & 1.76\rule{0em}{2.5ex} \\
& & \rule{0ex}{2.5ex}
expt. & 0.94 & 1.12 & 1.19 &  1.31 & $1.23$ & $1.39$ & $1.53$ & 1.67\\\hline
$P=+$ & n=1 & \rule{0ex}{2.5ex}
DSE & $1.82$ & $1.89$ & $1.94$ & $2.01$ &
$1.84$ & $1.95$ & $2.05$ & $2.14$\rule{0em}{2.5ex} \\
& & \rule{0ex}{2.5ex}
expt. & $1.44$ & $1.51$ & $1.66$ &  - & $1.60$ & - & - & -\\\hline
$P=-$ & n=0 & \rule{0ex}{2.5ex}
DSE & 1.82 & 1.92 & 1.96 & 2.04 & 2.07 & 2.16 & 2.26 & 2.36\rule{0em}{2.5ex} \\
& & \rule{0ex}{2.5ex}
expt. & $1.54$ & $1.67$ & $1.75$ & - & $1.65$ & $1.67$ & $1.82$ & - \\\hline
$P=-$ & n=1 & \rule{0ex}{2.5ex}
DSE & $1.89$ & $1.99$ & $2.02$ & $2.10$ & $2.11$ & $2.24$ & $2.33$ & $2.40$\rule{0em}{2.5ex} \\
& & \rule{0ex}{2.5ex}
expt. & $1.65$ & $1.80$ & - &  - & - & $1.94$ & - & -\\\hline
\end{tabular*}
\end{center}
\end{table*}

\begin{figure*}[t!]
\begin{centering}
\includegraphics[clip,width=0.46\textwidth]{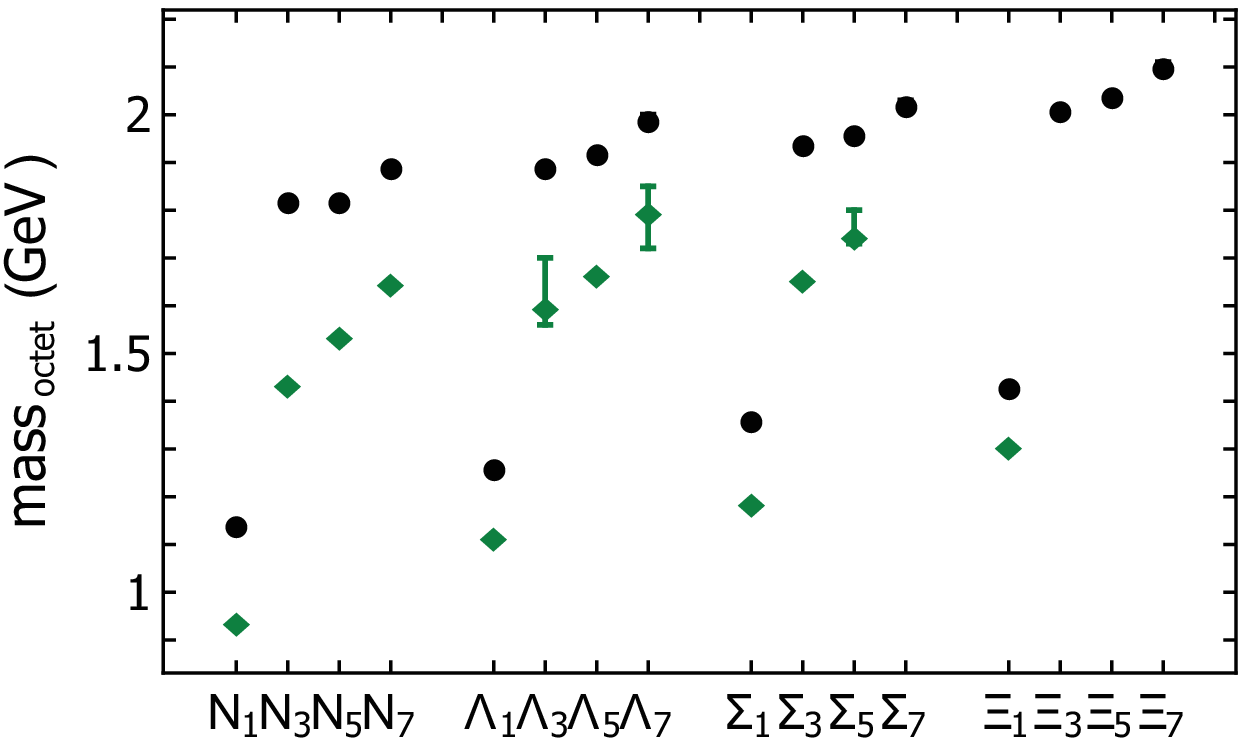}\hspace*{2em}
\includegraphics[clip,width=0.46\textwidth]{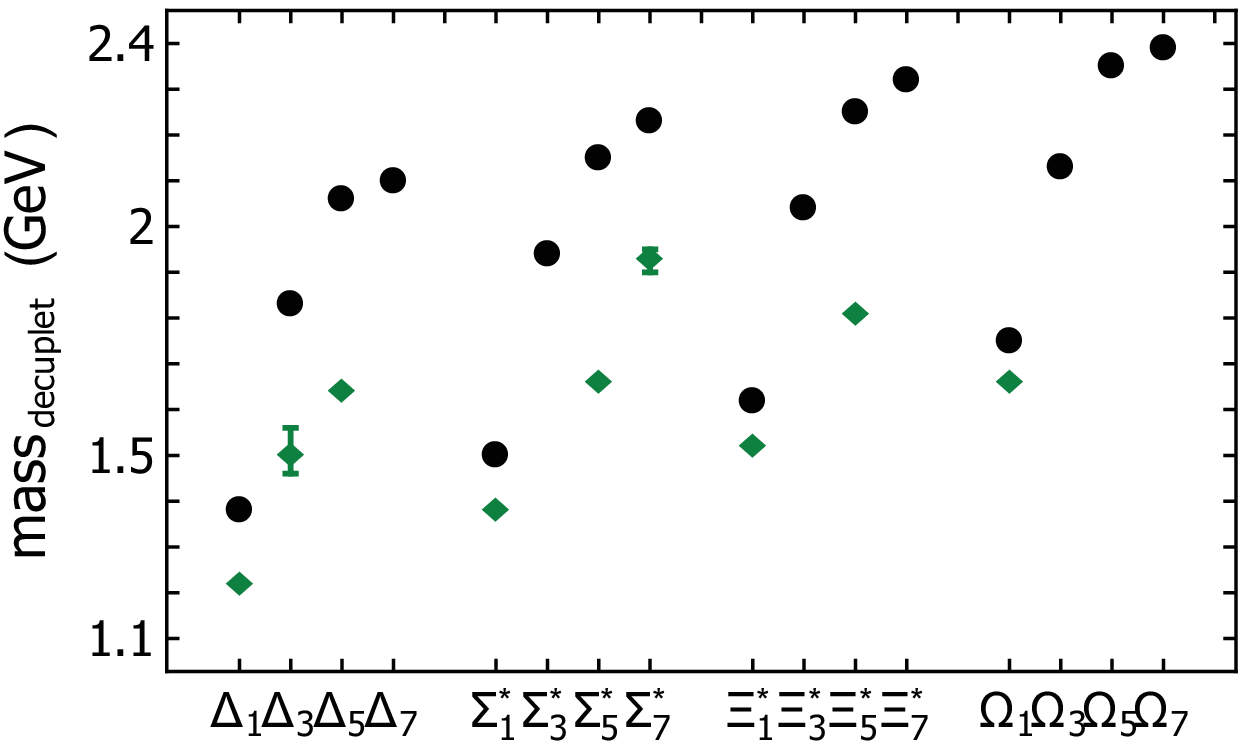}
\end{centering}
\caption{\label{PlotOctetDecupletMasses}
\underline{Left panel}: Pictorial representation of octet masses in Table~\protect\ref{OctetDecupletMasses}.  \emph{Circles} (black) -- computed masses; and \emph{diamonds} (green) -- empirical masses.  On the horizontal axis we list a particle name with a subscript that indicates its row in the table, \emph{e.g}.\ $N_1$ means nucleon column, row 1.  In this way the labels step through ground-state, radial excitation, parity partner, parity partner's radial excitation.
\underline{Right panel}: Analogous plot for the decuplet masses in Table~\protect\ref{OctetDecupletMasses}.
Where noticeable, the estimated uncertainty in the location of a resonance's pole position is indicated by an error bar.}
\end{figure*}

In the phenomenological application of a contact interaction, however, this difficulty has been skirted by means of an expedient employed in Refs.\,\cite{Volkov:1996br,Volkov:1999xf}, \emph{i.e}.\ one inserts a zero by hand into the kernels described above.   Plainly, the presence of this zero reduces the coupling in the Faddeev equation and hence increases the bound-state's mass.  Although this may not be as transparent with a more sophisticated interaction, a qualitatively equivalent mechanism is responsible for the elevated values of the masses of radial excitations \cite{Holl:2004fr, Qin:2011xq, Segovia:2015hra}.  We follow Refs.\,\cite{Roberts:2011cf, Chen:2012qr} in implementing this idea, \emph{viz}.\  the following replacement is made in each Faddeev kernel:
\begin{equation}
\label{groundtoradial}
\overline{\cal C}_1(\sigma) \to \overline{\cal F}_1(\sigma)\,,
\end{equation}
where $ \overline{\cal F}_1(\sigma) = \overline{\cal C}_1(\sigma) - d_{\cal F} \overline{\cal D}_1(\sigma)$,
\begin{equation}
\overline{\cal D}_1(\sigma)  =  \int_{r_{\rm uv}^2}^{r_{\rm ir}^2} d\tau\, \frac{2}{\tau^2} \,
\exp\left[-\tau \sigma \right].
\end{equation}
%
We use $1/d_{\cal F} = 2M_0^2$, with $M_0$ defined in Table~\ref{tabledressedquark}: a 20\% change in this value changes the mass of no radial excitation by more than 5\% \cite{Roberts:2011cf, Chen:2012qr}.

As already noted, a complete explanation of this method for analysing hadron radial excitations within a contact-interaction framework can be found elsewhere \cite{Roberts:2011cf, Chen:2012qr}.  The tool thus defined appears to produce sensible estimates for the masses of such baryons.  However, the Faddeev amplitudes are unrealistic, \emph{e.g}.\ the nucleon's radial excitation -- the Roper resonance -- is predicted to contain essentially no scalar diquark component, whereas realistic, momentum-dependent kernels predict that the scalar diquark content of both the nucleon and Roper is roughly 60\% \cite{Segovia:2015hra}.

\section{Baryon spectrum}\label{sectionthree}
%
One is now in a position to compute masses and Faddeev amplitudes for ground-state octet and decuplet baryons, their parity partners, and the first radial excitation of each of these systems.

\begin{figure*}[t!]
\begin{centering}
\includegraphics[clip,width=0.8\textwidth]{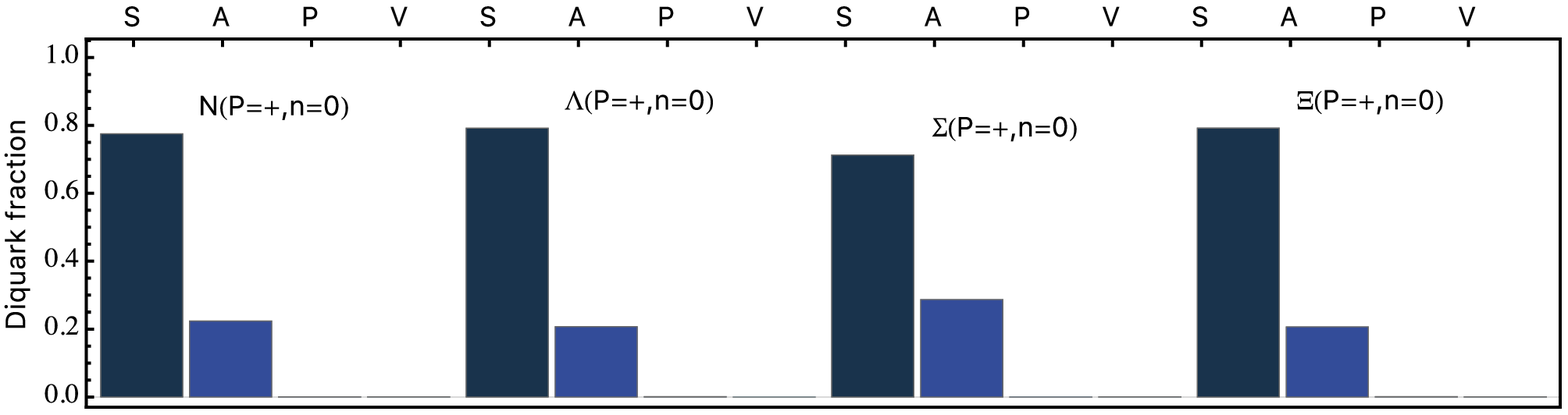}
\vspace*{-2ex}

\includegraphics[clip,width=0.8\textwidth]{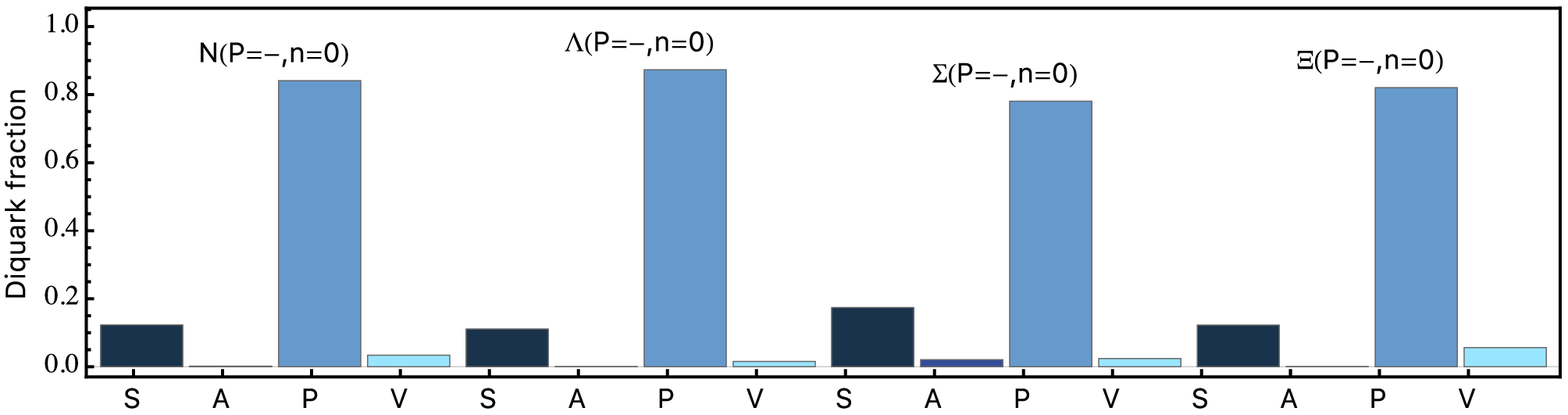}
\end{centering}
\caption{\label{diquarkprobabilities} Diquark content of octet baryons, computed from the unit-normalised Faddeev amplitudes listed in Table~\ref{FaddeevAmps}: \emph{upper panel}, ground-state even-parity systems; and \emph{lower panel}, their parity partners.}
\end{figure*}

In retracing the analysis of Ref.\,\cite{Chen:2012qr}, we have introduced two free parameters, \emph{viz}.\
$\mathpzc{s}_{\rm SO}$ in Eq.\,\eqref{eqsSO} and
$g_{\rm DB}$, described in connection with Eqs.\,\eqref{eqgDB1}, \eqref{eqgDB2}.
They are chosen in order to ensure that the dressed-quark core of the nucleon's parity partner is not lighter than that of its first radial excitation and the Faddeev amplitude of the $\Xi$ baryon's parity partner is dominated by negative-parity diquarks.  These complementary requirements are met with $\mathpzc{s}_{\rm SO}=1.8$, Eq.\,\eqref{eqsSO}, and
\begin{equation}
\label{ValuegDB}
g_{\rm DB}=0.1\,,
\end{equation}
values that yield the spectrum in Table~\ref{OctetDecupletMasses}, which is represented pictorially in Fig.\,\ref{PlotOctetDecupletMasses}.  Since changes of $\pm 10$\% in either parameter have no material impact on these results, we do not list a model error.  This low level of sensitivity to model-defining parameters enables us to achieve our primary aim of developing qualitative insights into the spectrum and structure of baryons, which can serve to inform a similarly wide-ranging analysis using realistic Faddeev kernels.  In this connection, it is notable that the value of $g_{\rm DB}$ is small and enforces significant suppression of unmatched-parity diquarks in a given $J=\tfrac{1}{2}$ bound-state: a value of $g_{\rm DB}=0.2$ would see positive-parity diquark correlations dominate in negative-parity octet baryons, thereby reducing their masses.  The issue of whether such suppression is realistic cannot be addressed until a sophisticated beyond-RL truncation is employed to treat an equally large array of baryon bound-states.

The Faddeev amplitudes associated with the masses listed in Table~\ref{OctetDecupletMasses} are catalogued in Table~\ref{FaddeevAmps}.  We consider that the $n=0$ (ground state) eigenvectors provide a sound guide; namely, a more realistic Faddeev equation kernel should produce roughly the same relative strengths of the various diquark components.  These strengths are depicted in Fig.\,\ref{diquarkprobabilities}, from which it is apparent that the parameter choices in Eqs.\,\eqref{eqsSO}, \eqref{ValuegDB} generate octet baryons with natural diquark content, \emph{i.e}.\ a baryon with parity ``P'' is constituted primarily from like-parity diquark correlations.

On the other hand, for the reasons detailed in connection with Eq.\,\eqref{groundtoradial}, the eigenvectors associated with a contact interaction treatment of radial excitations ($n=1$) must be treated with caution.  For instance, the Table~\ref{OctetDecupletMasses} prediction that the nucleon's radial excitation is almost purely constituted from axial-vector diquark correlations is now known to be an artefact of the contact interaction: using a QCD-kindred kernel, the scalar diquark component of this state is almost precisely the same as that of the ground-state nucleon \cite{Segovia:2015hra}.

\begin{figure}[t]
\begin{centering}
\includegraphics[clip,width=0.46\textwidth]{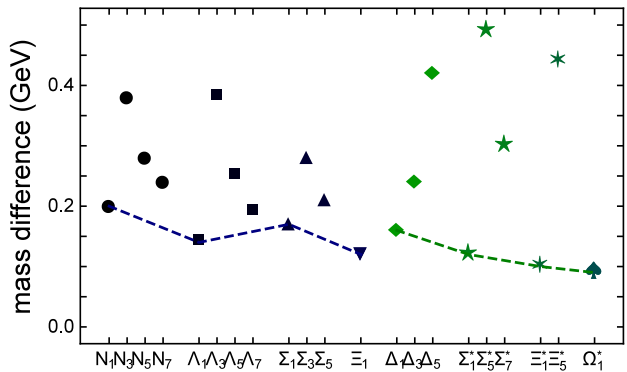}
\end{centering}
\caption{\label{masssplittings}
Theory-experiment mass differences computed, where possible, from Table~\protect\ref{OctetDecupletMasses}.
Horizontal axis: particle name with a subscript that indicates its row in the table, \emph{e.g}.\ $N_1$ means nucleon column, row\,1. }
\end{figure}

In Fig.\,\ref{masssplittings} we depict the theory-experiment mass-differences computed from Table~\ref{OctetDecupletMasses}.
This difference is uniformly less-than 0.2\,GeV for ground-states.  Moreover, it usually decreases as the number of $s$-quarks in the system is increased, as illustrated by the dashed lines in Fig.\,\ref{masssplittings}.
For octet systems, there is another regularity, \emph{viz}.\ the theory-experiment difference is greatest for the positive-parity radial excitations ($\approx 0.35\,$GeV), less for the $(P=-,n=0)$ states ($\approx 0.25\,$GeV), and least for the $(P=-,n=1)$ systems ($\approx 0.22\,$GeV).
The $(P=+,n=0)$, \emph{i.e}.\ ground-state, and $(P=+,n=1)$  differences are consistent with those found in  Ref.\,\cite{Chen:2012qr}, which ignored negative-parity diquark correlations in positive-parity baryons.  On the other hand, the results for parity partners are a marked contradiction, being much smaller herein.  It follows that no study of odd-parity baryons can be reliable unless it includes odd-parity diquark correlations.

Having made these observations, it is worth returning to the issue of meson-cloud effects on baryon masses.  In this connection, recall that our predictions, Table~\ref{OctetDecupletMasses}, are those for the mass of a given baryon's dressed-quark core, whereas the empirical values include effects associated with meson-baryon final-state interactions, which typically produce sizeable reductions \cite{Suzuki:2009nj, Gasparyan:2003fp}.  This was explained and illustrated for the nucleon and $\Delta$-resonance in Sect.\,4.5 of Ref.\,\cite{Roberts:2011cf} and in particular for the Roper resonance in Ref.\,\cite{Wilson:2011aa}.  Here we reiterate those instances in which a comparison can be made:
\begin{equation}
\rule{-0.9em}{0ex}
\begin{array}{l|llllll}
    & N_{940}^{P_{11}} & N_{1440}^{P_{11}} & N_{1535}^{S_{11}} & N_{1650}^{S_{11}}  & \Delta_{1232}^{P_{33}} & \Delta_{1700}^{D_{33}} \\\hline
{\rm herein} & 1.14 & 1.82 & 1.82 & 1.89 & 1.39 & 2.07
\\
M_{B}^0 \; \mbox{\cite{Suzuki:2009nj}}  & & 1.76 & 1.80 & 1.88 & 1.39 & 1.98\\
M_{B}^0 \; \mbox{\cite{Gasparyan:2003fp}} & 1.24 &  & 2.05 & 1.92 & 1.46 & 2.25
\end{array},
\end{equation}
where $M_B^0$, when it appears, is the relevant bare mass inferred in the associated dynamical coupled-channels (DCC) analysis \cite{Suzuki:2009nj, Gasparyan:2003fp}.\footnote{It is notable that the bare masses determined in Ref.\,\cite{Suzuki:2009nj} do not exceed our computed dressed-quark-core masses, whereas those inferred in Ref.\,\cite{Gasparyan:2003fp} are uniformly greater: relative error $=-7\pm 4$\%.  Since meson-baryon final-state interactions normally act to lower the pole-mass of a resonance, this observation suggests that Ref.\,\cite{Suzuki:2009nj} provides the more reliable indication of hadron quark-core masses.}
These bare masses have hitherto been uncertain and dependent on model details.  However, as we made no attempt to fit the $M_{B}^0$ values, their proximity to our results suggests that it might now be possible to place these bare masses on firmer ground, investing them with meaning within the context of hadron structure calculations that have a traceable connection with QCD.  Such a possibility is made more plausible by the fact that, with the inclusion of odd-parity diquark correlations, the agreement between our contact-interaction predictions and the DCC bare masses is much better than previously found, \emph{e.g}.\ whereas the mean relative-difference between contact-interaction core-masses and the bare masses from Ref.\,\cite{Suzuki:2009nj} was $15\pm 13$\% in the absence of odd-parity diquarks, it is herein just $2\pm 2$\%.

\section{Conclusion}\label{sectionconclusion}
Contemporary studies suggest that QCD possesses an infrared fixed point and therefore exhibits nearly conformal behaviour at infrared momenta \cite{Binosi:2016xxu, Binosi:2016nme, Aguilar:2002tc, Brodsky:2008be, Deur:2016tte}.  As a consequence, when defined and used judiciously, a vector$\,\times\,$vector contact interaction can be a useful tool for the analysis of numerous low-energy hadron properties.  Hadron masses are amongst these quantities, and we therefore used such an interaction to formulate Faddeev equations for the ground-state octet and decuplet baryons, their first radial excitation, and the parity partners of these systems.  The solutions of the equations yield the masses of the dressed-quark cores in these systems and information about the associated internal structure.  Notably, dynamical chiral symmetry breaking (DCSB) drives the formation of diquark correlations within baryons; and our Faddeev equation kernels incorporate all diquarks supported by a rainbow-ladder-like truncation of the equations describing such correlations, \emph{viz}.\ flavour-antitriplet scalar, pseudoscalar and vector, and flavour-sextet axial-vector.

A defect of rainbow-ladder (RL) truncation is that it suppresses angular momentum within hadron bound states.  This weakness typically leads to a poor description of all but the lowest-mass entries in the spectra of various hadron species.  Therefore, in formulating the bound-state equations relevant to a description of the baryon spectrum, we elaborated upon RL truncation by introducing phenomenological parameters that can ameliorate this sort of defect.  We thereby arrived at a model capable of yielding realistic insights, which can be used to guide subsequent studies with more sophisticated interactions.

Our key modification was the introduction of two parameters that work to express spin-orbit repulsion, one in the Bethe-Salpeter equations describing odd-parity diquarks and another in the Faddeev equations.  These two parameters, which find their origin in DCSB, were tuned in order to ensure: (\emph{i}) that the dressed-quark core of the nucleon's parity partner is not lighter than that of its first radial excitation; and (\emph{ii}) the Faddeev amplitude of the $\Xi$ baryon's parity partner is dominated by negative-parity diquarks.  The parameter values required to achieve these ends indicate that orbital angular momentum must play a significant role in the rest-frame wave functions of all so-called $P$-wave baryons.  Hence, a primary conclusion of our analysis is that DCSB is the key to explaining the mass splitting between parity partners and it works by enhancing repulsive spin-orbit interactions between all quarks in the bound state.

With this reformulation of the contact-interaction bound-state equations, we calculated the mass and Faddeev amplitudes of the dressed-quark cores in thirty-two isospin$+$strangeness-distinct baryons, \emph{i.e}.\ all octet and decuplet baryons, their lowest radial excitations, and the parity partners of all these systems.  The computed masses of all systems are reasonable, and the Faddeev amplitudes for ground states and their parity partners describe a realistic picture of the internal composition of these systems: ground-state, even parity baryons are constituted, almost exclusively, from like-parity diquark correlations, but odd-parity baryons must contain odd-parity diquarks, and such correlations should be dominant therein.

An important next step is to compute the spectrum studied herein using more realistic Faddeev equation kernels, \emph{e.g}.\ those built from the momentum-dependent propagators and diquark amplitudes used successfully in the description and prediction of nucleon elastic and nucleon-to-resonance transition form factors on a large domain of spacelike mometa \cite{Segovia:2014aza, Roberts:2015dea, Segovia:2015hra}.  This will help reveal just how realistic are the perspectives suggested by the contact-interaction analysis described herein.

\acknowledgments
We are grateful for constructive comments from Z.-F.~Cui, B.~El-Bennich, R.~Gothe, G.~Krein, C.~Mezrag, V.~Mokeev, S.-X. Qin, S.\,M.~Schmidt, C.~Shi and F.~Wang.
Work supported by:
National Natural Science Foundation of China (contract nos.\ 11475085, 11535005 and 11690030);
Funda\c{c}\~ao de Amparo \`a Pesquisa do Estado de S\~ao Paulo - FAPESP Grant No. 2015/21550-4;
U.S.\ Department of Energy, Office of Science, Office of Nuclear Physics, under contract no.~DE-AC02-06CH11357;
Chinese Ministry of Education, under the \emph{International Distinguished Professor} programme;
and the Alexander von Humboldt Foundation.

\appendix
\setcounter{equation}{0}
\setcounter{table}{0}
\renewcommand{\theequation}{\Alph{section}.\arabic{equation}}
\renewcommand{\thetable}{\Alph{section}.\arabic{table}}

\section{Assorted Formulae}
\label{App1}
We assume isospin symmetry throughout, in which case it is sufficient to specify the following spin-flavour column vectors for the octet baryons supported by our formulation of the contact-interaction:
{\allowdisplaybreaks
\begin{subequations}
\label{FaddeevSpinFlavour}
\begin{align}
u_p  = & \left[\begin{array}{c}
u[ud]_{0^+} \\ d\{uu\}_{1^+} \\ u\{ud\}_{1^+} \\ u[ud]_{0^-} \\ u[ud]_{1^-} \end{array}\right]
\leftrightarrow\left[\begin{array}{c}
\mathpzc{s}^1_p \\ \mathpzc{a}^4_p \\ \mathpzc{a}^5_p \\ \mathpzc{p}^1_p \\ \mathpzc{v}^1_p \end{array}\right]\,,
\\
u_\Lambda =\frac{1}{\sqrt{2}} & \left[\begin{array}{c}
\surd 2\, s[ud]_{0^+} \\ d[us]_{0^+}-u[ds]_{0^+} \\ d\{us\}_{1^+}-u\{ds\}_{1^+} \\
\surd 2\, s[ud]_{0^-} \\ d[us]_{0^-}-u[ds]_{0^-} \\ d\{us\}_{1^-}-u\{ds\}_{1^-} \end{array}\right]
\leftrightarrow\left[\begin{array}{c}
\mathpzc{s}^1_\Lambda \\ \mathpzc{s}^{[2,3]}_\Lambda \\ \mathpzc{a}^{[6,8]}_\Lambda \\
\mathpzc{p}^1_\Lambda \\ \mathpzc{p}^{[2,3]}_\Lambda \\ \mathpzc{v}^{[6,8]}_\Lambda \end{array}\right]\label{flavourLambda}\\
u_\Sigma = & \left[\begin{array}{c}
u[us]_{0^+} \\ s\{uu\}_{1^+} \\ u\{us\}_{1^+} \\ u[us]_{0^-} \\ u[us]_{1^-} \end{array}\right]
\leftrightarrow\left[\begin{array}{c}
\mathpzc{s}^2_\Sigma \\ \mathpzc{a}^4_\Sigma \\ \mathpzc{a}^6_\Sigma \\ \mathpzc{p}^2_\Sigma \\ \mathpzc{v}^2_\Sigma \end{array}\right]\,, \label{flavourSigma}
\\
u_\Xi = & \left[\begin{array}{c}
s[us]_{0^+} \\ s\{us\}_{1^+} \\ u\{ss\}_{1^+} \\ s[us]_{0^-} \\ s[us]_{1^-} \end{array}\right]
\leftrightarrow\left[\begin{array}{c}
\mathpzc{s}^2_\Sigma \\ \mathpzc{a}^6_\Sigma \\ \mathpzc{a}^9_\Sigma \\ \mathpzc{p}^2_\Sigma \\ \mathpzc{v}^2_\Sigma \end{array}\right]\,,
\end{align}
\end{subequations}}
\hspace*{-0.7\parindent}where $[\cdot]_{J^P}$ and $\{\cdot \}_{J^P}$ denote, respectively, flavour combinations generated by $T_{\bar 3_f}$ and $T_{6_f}$ in Eqs.\,\eqref{Tmatrices}, with the subscript indicating the spin-parity of the associated correlation.  Naturally, the same vector applies to ground-states, their parity partners, and the associated radial excitations.  The difference between these states is expressed in the values of the coefficients $\mathpzc{s}$, $\mathpzc{a}_{1,2}$, $\mathpzc{p}$, $\mathpzc{v}_{1,2}$ that appear in Eq.\,\eqref{spav} and are obtained by solving the appropriate Faddeev equations.  A shorthand notation for these coefficients, which expresses their connection with Eqs.\,\eqref{Tmatrices}, is specified by the rightmost column of each of Eqs.\,\eqref{FaddeevSpinFlavour}: superscript ``1'' connects with $T^1_{\bar 3_f}$, \ldots\,, superscript ``4''\,$\to T^1_{6_f}$, \ldots, superscript ``9''\,$\to T^6_{6_f}$.  The unit-normalised amplitudes obtained as solutions of our octet baryon Faddeev equations are listed in Table~\ref{FaddeevAmps}.

\begin{table*}[t]
\caption{\label{FaddeevAmps}
Contact-interaction Faddeev amplitudes for each of the octet baryons and their low-lying excitations.  The superscript in the expression $s^i$ or $a^i$ is a diquark enumeration label associated with Eq.\,\protect\eqref{FaddeevSpinFlavour}.
The rightmost column lists the square of the largest $J=0$ contribution to the amplitude, \emph{i.e}.\ it measures either the probability of finding a scalar diquark in an even-parity baryon or a pseudoscalar diquark in an odd-parity baryon.
N.B.\ In the notation of Eq.\,\eqref{spav}, we list only $\mathpzc{a}^4_{1,2}$ for the nucleon because $\mathpzc{a}_{1,2}^5= -\mathpzc{a}^4_{1,2}/\surd 2$.}
\begin{center}
\begin{tabular*}
{\hsize}
{
l@{\extracolsep{0ptplus1fil}}
l@{\extracolsep{0ptplus1fil}}
|r@{\extracolsep{0ptplus1fil}}
r@{\extracolsep{0ptplus1fil}}
r@{\extracolsep{0ptplus1fil}}
r@{\extracolsep{0ptplus1fil}}
r@{\extracolsep{0ptplus1fil}}
r@{\extracolsep{0ptplus1fil}}
r@{\extracolsep{0ptplus1fil}}
r@{\extracolsep{0ptplus1fil}}
r@{\extracolsep{0ptplus1fil}}
r@{\extracolsep{0ptplus1fil}}
r@{\extracolsep{0ptplus1fil}}
r@{\extracolsep{0ptplus1fil}}
r@{\extracolsep{0ptplus1fil}}
r@{\extracolsep{0ptplus1fil}}
r@{\extracolsep{0ptplus1fil}}
r@{\extracolsep{0ptplus1fil}}
r@{\extracolsep{0ptplus1fil}}
r@{\extracolsep{0ptplus1fil}}
r@{\extracolsep{0ptplus1fil}}
r@{\extracolsep{0ptplus1fil}}
|r@{\extracolsep{0ptplus1fil}}
}\hline
$(P,n)$
& & $\mathpzc{s}^1$ & $\mathpzc{s}^2$ & $\mathpzc{s}^{[2,3]}$ &
    $\mathpzc{a}^4_1$ & $\mathpzc{a}^4_2$ & $\mathpzc{a}_1^6$ & $\mathpzc{a}_2^6$ &
    $\mathpzc{a}_1^{[6,8]}$ & $\mathpzc{a}_2^{[6,8]}$ & $\mathpzc{a}_1^9$ & $\mathpzc{a}_1^9$ &
   $\mathpzc{p}^1$ & $\mathpzc{p}^2$ & $\mathpzc{p}^{[2,3]}$
   & $\mathpzc{v}_1^1$ & $\mathpzc{v}_2^1$ & $\mathpzc{v}_1^2$ & $\mathpzc{v}_2^2$ & $\mathpzc{v}_1^{[6,8]}$ &$\mathpzc{v}_2^{[6,8]}$ &
   $P_{J=0} $ \\\hline
%
$(+,0)$ \rule{0ex}{2.5ex}
& $N$       & 0.88 & & & 0.38 & -0.06 & & & & & & & 0.02 &  & & 0.02  & 0.00 & & & & & 77\%\\
\rule{0ex}{2.5ex} 
& $\Lambda$ & 0.66 & & 0.59 & & & & & 0.45 & 0.08 & & & 0.02  & & 0.03 & & & & & 0.01 & 0.00 & 79\%\\
\rule{0ex}{2.5ex} 
& $\Sigma$  & & 0.85 & & 0.45 & -0.26 & -0.13 & -0.01 & &  & & & & 0.01 & & & & 0.01 & 0.00 & & &  72\%\\
\rule{0ex}{2.5ex} 
& $\Xi$     & & 0.89 & &  & & -0.33 & 0.31 & & & -0.05 & -0.04 & & 0.03 &  & & & 0.02 & 0.00 & & & 79\%\\\hline
$(+,1)$ \rule{0ex}{2.5ex}
& $N$       & 0.02 & & & 0.52 & -0.62 & & & & & & & 0.07 &  & & 0.02  & -0.02 & & & & & 0\%\\
\rule{0ex}{2.5ex} 
& $\Lambda$ & 0.03 & & 0.06 & & & & & 0.77 & -0.62 & & & 0.14  & & 0.06 & & & & & 0.01 & -0.01 & 0\%\\
\rule{0ex}{2.5ex} 
& $\Sigma$  & & 0.02 & & 0.52 & -0.13 & -0.74 & 0.13 & &  & & & & 0.36 & & & & 0.13 & -0.11 & & &  0\%\\
\rule{0ex}{2.5ex} 
& $\Xi$     & & 0.03 & &  & & -0.31 & 0.42 & & & 0.15 & -0.32 & & 0.70 &  & & & 0.24 & -0.24 & & & 79\%\\\hline
$(-,0)$ \rule{0ex}{2.5ex}
& $N$       & 0.35 & & & 0.04 & 0.00 & & & & & & & 0.92 &  & & -0.05  & 0.18 & & & & & 84\%\\
\rule{0ex}{2.5ex} 
& $\Lambda$ & 0.25 & & 0.22 & & & & & 0.01 & -0.02 & & & 0.69  & & 0.63 & & & & & -0.05 & 0.12 & 87\%\\
\rule{0ex}{2.5ex} 
& $\Sigma$  & & 0.42 & & 0.11 & -0.06 & 0.06 & -0.03 & &  & & & & 0.88 & & & & 0.00 & 0.16 & & &  78\%\\
\rule{0ex}{2.5ex} 
& $\Xi$     & & 0.35 & &  & & -0.01 & 0.01 & & & 0.01 & -0.02 & & 0.91 &  & & & -0.11 & 0.21 & & & 82\%\\\hline
$(-,1)$ \rule{0ex}{2.5ex}
& $N$       & 0.53 & & & 0.28 & 0.26 & & & & & & & 0.62 &  & & 0.34  & 0.04 & & & & & 39\%\\
\rule{0ex}{2.5ex} 
& $\Lambda$ & 0.41 & & 0.37 & & & & & 0.28 & 0.25 & & & 0.36  & & 0.60 & & & & & 0.25 & -0.05 & 49\%\\
\rule{0ex}{2.5ex} 
& $\Sigma$  & & 0.52 & & 0.37 & -0.18 & 0.39 & -0.17 & &  & & & & 0.56 & & & & 0.27 & 0.02 & & &  31\%\\
\rule{0ex}{2.5ex} 
& $\Xi$     & & 0.61 & &  & & -0.21 & 0.22 & & & -0.19 & 0.22 & & 0.58 &  & & & 0.34 & 0.04 & & & 34\%\\\hline
\end{tabular*}
\end{center}
\end{table*}

It is worth comparing Eqs.\,\eqref{flavourLambda} and \eqref{flavourSigma}, with the latter adapted to the neutral $\Sigma^0$ case following Eq.\,(49) in Ref.\,\cite{Chen:2012qr}.  Whilst the $\Lambda^0$ and $\Sigma^0$ baryons are associated with the same combination of valence-quarks,
their spin-flavour wave functions are different: the $\Lambda_{I=0}^0$ contains more of the lighter $J=0$ diquark correlations than the $\Sigma_{I=1}^0$.  It follows that the $\Lambda^0$ must be lighter than the $\Sigma^0$.  The mechanism underlying this splitting is analogous to that which produces the $\pi$-$\rho$ mass difference, and also to that associated with the colour-hyperfine interaction used in quark models.  It is realised here via the breaking of isospin-symmetry in the associated baryon wave functions.

The analogous vectors for the decuplet baryons are:
{\allowdisplaybreaks
\begin{subequations}
\label{DecupletSpinFlavour}
\begin{align}
u_\Delta = & \left[ u \{uu\}_1^+ \right] \leftrightarrow [f^4_\Delta]\\
u_{\Sigma^\ast} = & \left[\begin{array}{c}
s\{uu\}_{1^+}\\ u \{us\}_{1^+}
\end{array}\right] \leftrightarrow
\left[\begin{array}{c}
f^4_{\Sigma^\ast} \\ f^6_{\Sigma^\ast}
\end{array}\right]\\
u_{\Xi^\ast} = & \left[\begin{array}{c}
s\{us\}_{1^+}\\ u \{ss\}_{1^+}
\end{array}\right] \leftrightarrow
\left[\begin{array}{c}
f^6_{\Xi^\ast} \\ f^9_{\Xi^\ast}
\end{array}\right]\\
u_\Omega = & \left[ s \{ss\}_1^+ \right] \leftrightarrow [f^9_\Delta]\,.
\end{align}
\end{subequations}}
Only the $\Sigma^\ast$ and $\Xi^\ast$ systems possess nontrivial unit-normalised Faddeev amplitudes:
\begin{equation}
\begin{array} {lc|rrr}
 (P,n) & & f^4 & f^6 & f^9\\\hline
(+,0) & \Sigma^\ast & 0.61 & 0.79  \\
        & \Xi^\ast & & 0.85 & 0.52 \\\hline
(+,1) & \Sigma^\ast & 0.71 & 0.71  \\
        & \Xi^\ast & & 0.88 & 0.47 \\\hline
(-,0) & \Sigma^\ast & 0.68 & 0.74  \\
        & \Xi^\ast & & 0.87 & 0.49 \\\hline
(-,1) & \Sigma^\ast & 0.67 & 0.75  \\
        & \Xi^\ast & & 0.87 & 0.50 \\\hline
\end{array}\,.
\end{equation}
In each case, the mixed-flavour diquark is favoured for an obvious reason, \emph{viz}.\ considering the Faddeev equation kernel, it is fed by twice as many exchange processes as the like-flavour correlation.

We do not record the baryon Faddeev equations herein.  They are readily derived, following the procedures detailed in Refs.\,\cite{Roberts:2011cf, Chen:2012qr}, but the final expressions for octet systems are lengthy with, \emph{e.g}.\ the kernel for the $\Sigma$-baryon involving sixty-four entries.  On the other hand, each such entry has a simple form, similar to that on the right-hand-side of Eq.\,\eqref{DeltaFE}, and all kernels are recognisable extensions of those listed in full elsewhere \cite{Roberts:2011cf, Chen:2012qr}.  Hence, those interested in repeating the calculations described herein need only ensure they can recover the elements already published and then complete the kernel matrices using the same methods.



\end{document}